\documentclass{elsart}
\usepackage{epsfig}

\begin{document}

\begin{frontmatter}

\title{Relating production and masses of the vector and 
{\mathversion{bold} $P$}-wave mesons
             for light and heavy flavours at LEP}

\author{P.V. Chliapnikov}

\address{Institute for High Energy Physics, Protvino, RU-142284, 
        Russia\thanksref{mail}}
\thanks[mail]{E-mail address: chliapnikov@mx.ihep.su 
                                             (P.V. Chliapnikov)}

\begin{abstract}
{\scriptsize
The production rates of primary vector and $P$-wave mesons 
in Z$^0$ hadronic decays are analysed. The mass dependence of 
production rates for the bottom, charm, strange charm and three
families of the light-flavour mesons is found to be very similar, 
allowing to relate the relative production rates for mesons with 
different flavours and, possibly, their masses. The strange axial 
mesons K$_1(1273)$ and K$_1(1402)$ might be assigned to the  
$1^+(1/2)$ and $1^+(3/2)$ levels degenerate with the $0^+(1/2)$ 
and $2^+(3/2)$ levels of the K$^*_0(1430)$ and K$^*_2(1430)$, 
respectively, if the observed K$^*_0(1430)$ mass is replaced by 
its ``bare'' $q\bar q$ mass corresponding to the $K$-matrix pole 
and close to the $K_1(1273)$ mass. Then the $0^+(1/2)$ and 
$1^+(1/2)$ levels are {\em below\/} the $1^+(3/2)$ and $2^+(3/2)$ 
levels for the strange, charm and bottom mesons. 
}
\end{abstract}

\end{frontmatter}

\def\zp{Z.\ Phys.\ {\bf C}}
\def\epj{Eur.\ Phys.\ J.\ {\bf C}}
\def\pl{Phys.\ Lett.\ {\bf B}}
\def\prl{Phys.\ Rev.\ Lett.\ }
\def\pr{Phys.\ Rev.\ {\bf D}}
\def\ac{ALEPH\ Collab.,\ }
\def\dca{DELPHI\ Collab.,\ P.\ Abreu\ et\ al.,\ }
\def\lc{L3\ Collab.,\ }
\def\oc{OPAL\ Collab.,\ }
\def\beq{\begin{equation}}
\def\eeq{\end{equation}}
\def\bear{\begin{eqnarray}}
\def\enar{\end{eqnarray}}
\def\nnb{\nonumber}
\def\nin{\noindent}
\def\la{\langle}
\def\ra{\rangle}
\def\ol{\overline}
%
\newcommand \zo {\ifmmode \mathrm{Z^0}    \else $\mathrm{Z^0}$\fi}
\newcommand \n  {\ifmmode \la n \ra       \else $\la n \ra$\fi}
%
%
\newcommand \kl{\ifmmode \mathrm{K}       \else $\mathrm{K}$\fi}
\newcommand \pis{\ifmmode \pi             \else $\pi$\fi}
\newcommand \dl{\ifmmode \mathrm{D}       \else $\mathrm{D}$\fi}
\newcommand \bl{\ifmmode \mathrm{B}       \else $\mathrm{B}$\fi}
%
%
\newcommand \roz{\ifmmode \rho^0          \else $\rho^0$\fi}
\newcommand \oms{\ifmmode \omega          \else $\omega$\fi}
\newcommand \ph{\ifmmode \phi             \else $\phi$\fi}
\newcommand \kvn{\ifmmode  \kl^*(892)     \else $\kl^*(892)$\fi}
\newcommand \kvo{\ifmmode  \kl^{*0}       \else $\kl^{*0}$\fi}
\newcommand \kvon{\ifmmode \kl^{*0}(892)  \else $\kl^{*0}(892)$\fi}
\newcommand \dv{\ifmmode   \dl^*          \else $\dl^*$\fi}
\newcommand \dvp{\ifmmode \mathrm{D^{*+}} \else $\mathrm{D^{*+}}$\fi}
\newcommand \dvs{\ifmmode   \dl^{*+}_s    \else $\dl^{*+}_s$\fi}
\newcommand \bv{\ifmmode \mathrm{B}^*     \else $\mathrm{B}^*$\fi}
%
%
\newcommand \ft{\ifmmode f_2              \else $f_2$\fi}
\newcommand \ftn{\ifmmode f_2(1275)       \else $f_2(1275)$\fi}
\newcommand \fpn{\ifmmode f'_2(1525)      \else $f'_2(1525)$\fi}
\newcommand \kto{\ifmmode \kl_2^{*0}      \else $\kl_2^{*0}$\fi}
\newcommand \ktn{\ifmmode \kl_2^*(1430)   \else $\kl_2^*(1430)$\fi}
\newcommand \kton{\ifmmode\kl_2^{*0}(1430)\else $\kl_2^{*0}(1430)$\fi}
\newcommand \ka{\ifmmode\kl_1             \else $\kl_1$\fi}
\newcommand \kal{\ifmmode\kl_1(1273)      \else $\kl_1(1273)$\fi}
\newcommand \kah{\ifmmode\kl_1(1402)      \else $\kl_1(1402)$\fi}
\newcommand \dt{\ifmmode  \dl^*_2         \else $\dl_2^*$\fi}
\newcommand \dto{\ifmmode \dl^{*0}_2      \else $\dl_2^{*0}$\fi}
\newcommand \dton{\ifmmode\dto(2460)      \else $\dto(2460)$\fi}
\newcommand \don{\ifmmode \dl^0_1(2420)   \else $\dl^0_1(2420)$\fi}
\newcommand \doz{\ifmmode \dl^*_1         \else $\dl^*_1$\fi}
\newcommand \dso{\ifmmode \dl^+_{s1}      \else $\dl^+_{s1}$\fi}
\newcommand \dson{\ifmmode\dl^+_{s1}(2536)\else $\dl^+_{s1}(2536)$\fi}
\newcommand \bo{\ifmmode   \bl_1          \else $\bl_1$\fi}
\newcommand \boz{\ifmmode  \bl^*_1        \else $\bl^*_1$\fi}
\newcommand \bt{\ifmmode   \bl_2^*        \else $\bl_2^*$\fi}
\newcommand \bzz{\ifmmode  \bl^{**}       \else $\bl^{**}$\fi}
%
\newcommand \fz{\ifmmode f_0              \else $f_0$\fi}
\newcommand \fzn{\ifmmode f_0(980)        \else $f_0(980)$\fi}
\newcommand \az{\ifmmode a_0              \else $a_0$\fi}
\newcommand \azn{\ifmmode a^+_0(980)      \else $a^+_0(980)$\fi}
\newcommand \kz{\ifmmode   \kl^*_0        \else $\kl^*_0$\fi}
\newcommand \kzn{\ifmmode  \kl^*_0(1430)  \else $\kl^*_0(1430)$\fi}
\newcommand \dz{\ifmmode \dl_0            \else $\dl_0$\fi}
\newcommand \bz{\ifmmode \bl_0            \else $\bl_0$\fi}

The LEP experiments accumulated rich information on inclusive 
production of the light-flavour, charm and bottom mesons in the 
\zo\ hadronic decays including data on $P$-wave meson production. 
In this Letter, we use these data to compare the production of 
primary  vector and $P$-wave mesons in an attempt to relate the 
production of these states for light and heavy flavours. 

The total production rates of the vector \roz, \oms, \kvon\ and 
\ph, the tensor \ftn, \kton\ and \fpn, and the scalar \fzn\ and 
\azn\ mesons measured by the LEP experiments 
\cite{Aleph,Drofzftkt,Dkvophi,Lom,Okvkt,Ophfzft,Oazom}
are presented in Table~1. For the vector and scalar mesons, the 
measurements from the different LEP experiments agree within
errors. Therefore subsequently we used the rates obtained by 
averaging the results of these experiments, also presented in
Table~1.
\renewcommand{\arraystretch}{1.}
\begin{table}[tphb]
\caption[] {\scriptsize The total production rates of the vector, 
tensor and scalar mesons in the light-quark sector measured by the 
LEP experiments, averaged total rates for the vector and scalar 
mesons, fractions of primary mesons obtained from the JETSET model 
and direct rates determined by multiplying the total rates by the 
fractions of primary mesons. For the \kvn, \ktn\ and \azn\ 
antiparticles and charge conjugates are not included into the 
definition of the rates.}
\begin{center}
\begin{tabular}{|lcccc|} \hline
Meson  &Total rate &Averaged rate &Fraction &Direct rate\\
\hline
\roz\  &$1.45\pm 0.21$     \cite{Aleph}     &$1.23\pm0.10$ 
                           &0.54  &$0.664\pm 0.054$ \\ 
       &$1.19\pm 0.10$     \cite{Drofzftkt} & & &\\
\oms   &$1.07\pm 0.14$     \cite{Aleph}     &$1.08\pm 0.09$ 
                           &0.57  &$0.618\pm 0.049$ \\
       &$1.17\pm 0.17$     \cite{Lom}       & & &\\
       &$1.04\pm 0.15$     \cite{Oazom}     & & & \\
\kvon\ &$0.415\pm 0.045$   \cite{Aleph}     &$0.377\pm 0.017$
                           &0.60  &$0.226\pm 0.010$\\
       &$0.385\pm 0.040$   \cite{Dkvophi}   & & &\\
       &$0.379\pm 0.017$   \cite{Okvkt}     & & &\\
\ph\   &$0.122\pm 0.009$   \cite{Aleph}     &$0.0966\pm 0.0073$
                           &0.70  &$0.0676\pm 0.0051$ \\
       &$0.104\pm 0.008$   \cite{Dkvophi}   & & &\\
       &$0.091\pm 0.004$   \cite{Ophfzft}   & & &\\
\hline 
\ftn\  &$0.155\pm 0.021$   \cite{Drofzftkt} & &0.96  
                           &$0.149\pm 0.020$\\
       &$0.214\pm 0.038$   \cite{Ophfzft}   & & & \\
\kton\ &$0.037\pm 0.013$   \cite{Drofzftkt} &
                           &0.98  &$0.036\pm 0.013$\\
       &$0.119\pm 0.044$   \cite{Okvkt}     & & & \\
\fpn\  &$0.012\pm 0.006$   \cite{Drofzftkt} & 
                           &0.98  &$0.012\pm 0.006$\\
\hline
\fzn\  &$0.164\pm 0.021$   \cite{Drofzftkt} &$0.147\pm 0.011$ 
                           &0.93  &$0.137\pm 0.010$\\
       &$0.141\pm 0.013$   \cite{Ophfzft}   & & & \\
\azn\  &$0.1350\pm 0.0055$ \cite{Oazom}     & 
                           &0.93  &$0.126\pm 0.051$\\
\hline                                             
\end{tabular}
\end{center}
\end{table}
In calculating the errors of averages, the standard procedure 
suggested by the PDG group \cite{PDG} was applied. The DELPHI 
\cite{Drofzftkt} and OPAL \cite{Okvkt,Ophfzft} results on the 
\ftn\ and \kton\ rates are less consistent. The \kton/\ftn\ ratio 
from DELPHI, $0.24\pm 0.09$, agrees with usually accepted value 
of the strangeness suppression parameter $\lambda \approx 0.3$. 
This is also true within large errors for the ratio $\fpn/\kton 
= 0.32 \pm 0.20$. The same ratios $\kton/\ftn  = 0.56\pm 0.23$ and 
$\fpn/\kton = 0.10 \pm 0.06$ obtained from the \ftn\ and \kton\ 
rates from OPAL and the \fpn\ rate from DELPHI differ from the 
expected value of $\lambda$ by a factor of 2 and 3, respectively, 
although consistent with it within 1 and 3 standard deviations. 
For this reason we subsequently relied on the DELPHI measurements 
of the \ftn\ and \kton\ total rates. The direct production rates 
were determined by multiplying the total averaged rates by the 
fractions of primary mesons obtained from the JETSET model 
\cite{jetset} given in \cite{pei} and reproduced in Table~1. 
The fractions of the promptly produced vector mesons are quite high. 
For the tensor mesons, they are close to 1. This facilitates the 
analysis of the vector and tensor meson direct rates in comparison 
with the more difficult situation for the pseudoscalar mesons 
\cite{h_2,compar}\footnote{\scriptsize Although the JETSET model 
predictions for the ratios of the promptly produced vector and 
pseudoscalar mesons are also quite compatible with experiment 
\cite{compar}.}.

The direct production rates of the vector and tensor mesons per 
spin projection, $\n/(2J+1)$, are also presented as a function 
of their mass, $M$, in Fig.~1. The mass dependences of the \roz, 
\oms\ and \ftn, the \kvon\ and \kton, the \ph\ and \fpn\ rates 
are very similar. The fit of the data to three exponentials
$\n/(2J+1) = a \mathrm{e}^{-b M}$, with different normalization 
parameters $a$ but the {\em same\/} slope parameter, yields 
$b = 4.11 \pm 0.27$ (GeV/$c^2$)$^{-1}$.
\begin{figure}[thbp]
\noindent
\begin{minipage}{0.50\linewidth}
\includegraphics[width=\linewidth]{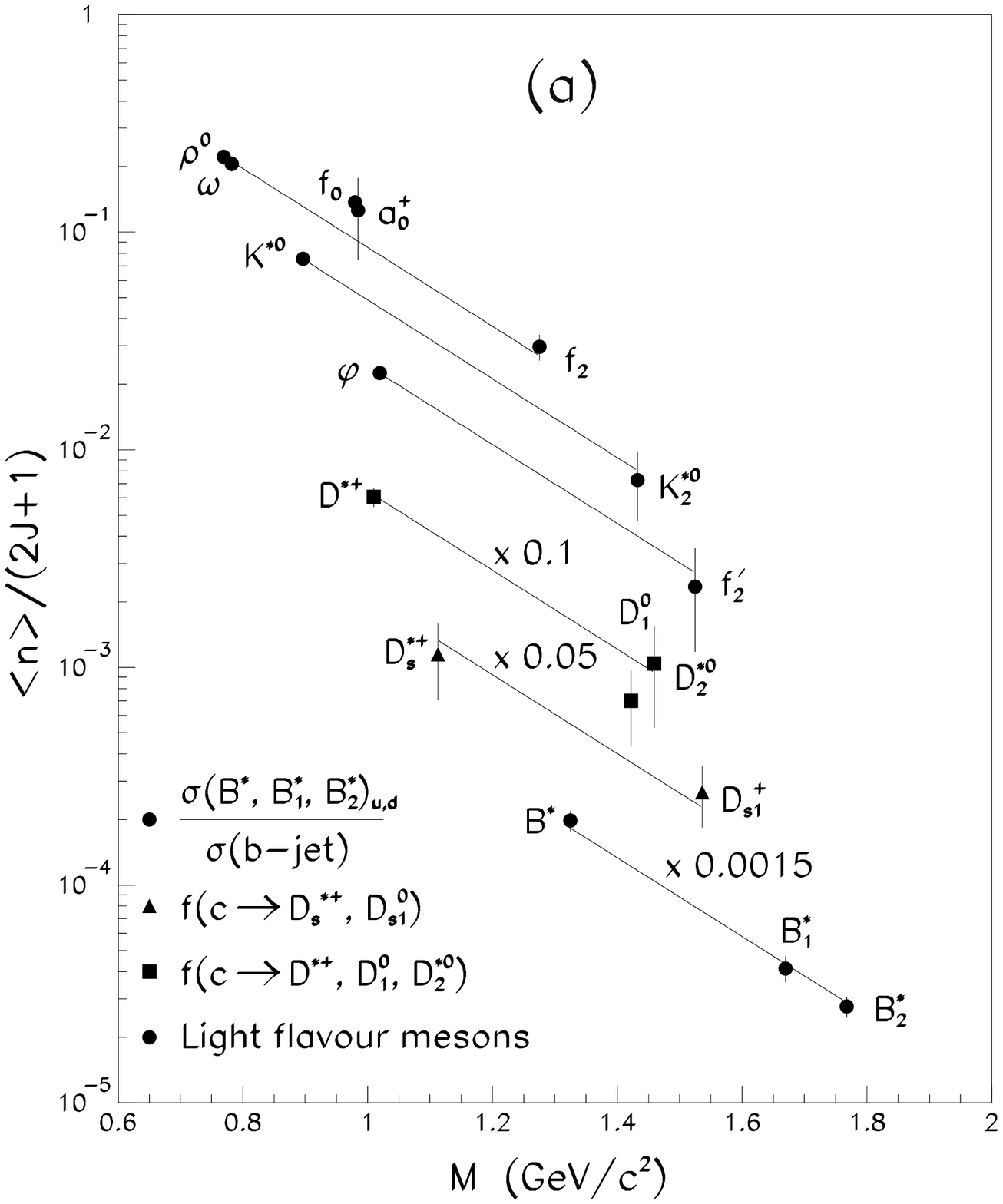}
\end{minipage}\hfill
\begin{minipage}{.50\linewidth}
\includegraphics[width=\linewidth]{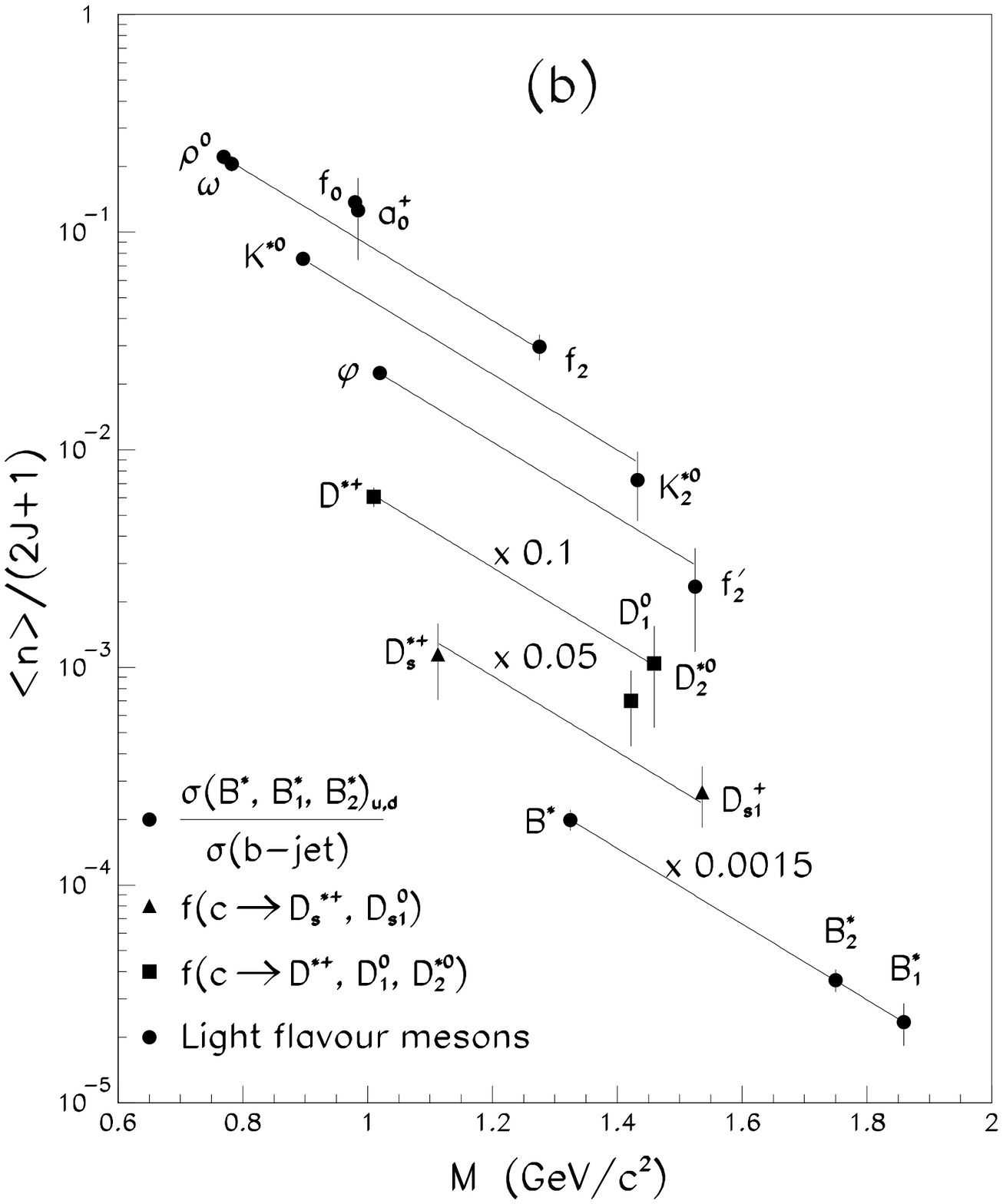}
\end{minipage}\hfill
   \caption{\scriptsize The mass dependence of direct production 
rates, \n, for light-flavour mesons, fragmentation fractions 
$f(c \rightarrow \dvp, \dl^0_1, \dto)$ and $f(c \rightarrow 
\dvs, \dso)$ for charm mesons, and ratios $\sigma 
(\bv, \boz, \bt)_{u,d}/\sigma_{b-\mathrm{jet}}$ for bottom mesons 
(with their masses and rates from L3 (a) and OPAL (b)), all 
divided by the spin counting factor $2J+1$. The data points for 
charm, strange charm and bottom mesons have been scaled by 
factors of 0.1, 0.05 and 0.0015 respectively and shifted by 1 
GeV/$c^2$ for charm and by 4 GeV/$c^2$ for bottom mesons for 
clarity. The solid lines represent the result of the fit of the 
data for the light-flavour vector and tensor mesons, and the 
charm and bottom vector and $P$-wave mesons to six exponentials 
with the {\em same\/} slope. }
\end{figure}

The probabilities that a charm quark fragments into the $P$-wave 
\dton, \don\ and \dson\  mesons were measured by OPAL \cite{Odt}: 
\bear
 f(c \rightarrow \dton) & = & 0.052 \pm 0.026\\ 
 f(c \rightarrow \don)  & = & 0.021 \pm 0.008 \\
 f(c \rightarrow \dson) & = & 0.016 \pm 0.005.
\enar
The charm fragmentation fraction into the \dvp\ meson measured 
by ALEPH \cite{Advp}, DELPHI \cite{Ddvp} and OPAL \cite{Odvp}  
amounted to $0.233\pm 0.015$, $0.255\pm 0.017$ and $0.222\pm 0.020$,
respectively. The averaged result is
\beq
     f(c \rightarrow \dvp) = 0.238 \pm 0.010.
\eeq
The charm fragmentation fraction into the \dvs\ measured by ALEPH 
\cite{Advp} is 
\beq
     f(c \rightarrow \dvs) = 0.069 \pm 0.026.
\eeq
Accounting for the $\dl^*_2(2460)$, $\dl_1(2420)$ and \dson\ decays 
into \dvp\pis\ and \dvp\kl, and assuming isospin invariance,
the charm fragmentation fraction into the primary \dvp\ meson is 
\beq
     f(c \rightarrow \dvp_{\mathrm{prompt}}) = 0.183 \pm 0.018.
\eeq
The \dvs\ in Eq.~(5) has been considered as promptly produced, 
since a contribution of possibly seen \dson\ decay into $\dvs\gamma$ 
can be ignored within presently large errors. The values of the 
charm fragmentation fractions into the primary vector mesons \dvp\ 
and \dvs\ and $P$-wave mesons \don, \dton\ and \dson, divided by 
the corresponding spin counting factors $2J +1$, are presented in 
Fig.~1. As one can see their mass dependence is very similar to the 
one observed for the light-flavour vector and tensor mesons.
  
The experimental situation for $P$-wave meson production in the 
bottom-quark sector is more complicated. In the quark model one 
expects for each spectator flavour four different orbitally excited 
states. For $b\bar u$ and $b\bar d$ states they are commonly 
labelled as $\bzz_{u,d}$. Heavy quark effective symmetry (HQET) 
\cite{HQET} groups these four states into two doublets with 
$j_q = 1/2$ and $j_q = 3/2$ where $\vec{j}_q = \vec{s}_q + \vec{l}$ 
is the total angular momentum of the light quark. The $j_q = 1/2$ 
doublet consists of the states $\bl_0$ and $\bl^*_1$ with spins 0 
and 1 respectively. The states $\bl_1$ and $\bl^*_2$, with respective 
spins 1 and 2, comprise the $j_q = 3/2$ doublet. The splitting 
between the states in each doublet is expected to be small. The 
states in the $j_q = 1/2$ doublet are expected to be broad since 
they can decay through an $S$-wave transition, whereas the 
$j_q =3/2$ states decay through a $D$-wave transition and are 
therefore thought to be narrow.

Evidence for the \bzz\ states with inclusively reconstructed 
\bl\ mesons has been clearly observed by the LEP experiments 
\cite{Abvbt,Dbt,Lbt,Obt}. The relative rate of all spin states, 
$\sigma_{\bzz_{u,d}}/\sigma_{b-\mathrm{jet}}$ amounted to 
$0.214 \pm 0.049$ \cite{Abvbt}, $0.270\pm 0.063$ \cite{Dbt}, 
$0.320\pm 0.067$ \cite{Lbt} and $0.270 \pm 0.056$ \cite{Obt}. 
Besides, ALEPH reported the rate of $0.238\pm 0.085$ \cite{Abtexcl} 
based on fully reconstructed \bl\ mesons. The cited values from 
ALEPH were obtained using the fraction of $\bl_{u,d}$ mesons in 
$\zo \rightarrow b\bar b$ decays, $0.768\pm 0.052$, taken from 
\cite{Abvbt}. Averaging these results we obtain 
\beq
\sigma_{\bzz_{u,d}}/\sigma_{b-\mathrm{jet}} = 0.258 \pm 0.027.
\eeq
A similar value, $0.28 \pm 0.06 \pm 0.03$, was recently measured
in $p\bar p$ collisions at $\sqrt s = 1.8$ TeV by CDF \cite{CDF}.

The relative production rates, masses and widths of the four 
different states contributing to the $\bzz_{u,d}$ signal are 
not very well known. In the framework of HQET, attempts have 
been made by ALEPH \cite{Abtexcl}, L3 \cite{Lbt} and OPAL 
\cite{Ob1nar} to determine the masses and widths of at least 
one of these states. The fitted masses are shown (as bold 
numbers) in Table 2. The complicated fitting procedures and 
constraints in these experiments were different, apart from 
mass splitting between the states belonging to the same $j_q$ 
doublet. For the narrow states, the constraint $M_{\bt}-M_{\bo} 
= 12$ MeV/$c^2$ was applied by all experiments. For the broad 
states, ALEPH and L3 applied the same constraint, while OPAL 
took $M_{\boz} - M_{\bz} = 20$ MeV/$c^2$. ALEPH fitted the mass 
of the \bt\ meson only, with the constraint $M_{\bt} - M_{\boz} 
= 100$ MeV/$c^2$. The masses of the states resulting from these 
constraints are also shown in Table~2. 
\renewcommand{\arraystretch}{1.}
\begin{table}[htp]
\caption[] {\scriptsize Masses of the $P$-wave bottom mesons 
           determined by the LEP experiments from the fits to 
           the data (bold numbers) together with masses of their 
           partners used as the constraints in the fits. The 
           branching fractions into \bv\pis\ used by ALEPH, L3 
           and in the present Letter are also shown.}
\begin{center}
\begin{tabular}{|cccccc|} \hline
Meson &$J^P_{j_q}$ &ALEPH \cite{Abtexcl}     &L3 \cite{Lbt} &OPAL 
                               \cite{Ob1nar} &Br(\bv\pis) \\
\hline
\bz  &$0^+_{1/2}$ &$5627\pm ^8_{11}\pm ^6_4$ &$5658\pm 10\pm 13$ 
          &{\mathversion{bold}$5839\pm ^{13}_{14} \pm^{34}_{42}$}   
                                                         &0.0 \\
\boz &$1^+_{1/2}$ &$5639\pm ^8_{11}\pm ^6_4$
                  &{\mathversion{bold}$5670\pm 10\pm 13$}
                  &$5859\pm ^{13}_{14}\pm ^{34}_{42}$    &1.0 \\
\bo  &$1^+_{3/2}$ &$5727\pm ^8_{11}\pm ^6_4$   &$5756\pm 5\pm 6$
                  &{\mathversion{bold}$5738\pm^5_6\pm 7$}&1.0 \\
\bt  &$2^+_{3/2}$ &{\mathversion{bold}$5739\pm ^8_{11}\pm ^6_4$} 
                  &{\mathversion{bold}$5768\pm 5\pm 6$} 
                  &$5750\pm^5_6\pm 7$                    &0.5 \\
\hline
\end{tabular}
\end{center}
\end{table}

The results on the masses of the narrow \bo\ and \bt\ are quite 
consistent bearing in mind the difference in other assumptions in 
the corresponding  fits. The \bo\ mass, $5710 \pm 20$ MeV/$c^2$, 
extracted by CDF \cite{CDF}, with the error not including the 
theoretical uncertainty on the shape of the \bzz\ peak, is also 
consistent with the LEP results. On the other hand, the masses of 
the broad \boz\ and \bz\ states obtained in the L3 fit and 
constrained by ALEPH are smaller by $\approx 200$ MeV/$c^2$ than 
the masses determined by OPAL, even if OPAL stressed that the \bz\ 
mass could not be considered as a robust fit result. Theoretical 
predictions (see \cite{Godfrey,Eichten,Gronau,Gupta,Isgur,Ebert} 
and references therein) for the masses of the four spin states in 
the charm and bottom sectors are also different. Some models 
\cite{Gronau,Gupta} predict that the broad $j_q = 1/2$ states 
have smaller masses than the narrow $j_q = 3/2$ states, in 
agreement with the L3 result and ALEPH constraints. Other models 
\cite{Isgur,Ebert} proposing spin-orbit inversion are more 
consistent with the OPAL result. The difference in the experimental 
results might be, at least partly, explained by different assumptions 
about the relative production rates of the four states. The 
corresponding proportions were set by ALEPH, L3 and CDF according 
to simple total spin counting, \bz:\boz:\bo:\bt  = 1:3:3:5. OPAL 
fixed the relative production rates of the same states to 2:2:3:3. 

Our attempt to determine the relative rates of the four spin states 
is based on the assumption that the mass dependence of their 
production rates is the same as that observed for the light-flavour 
and charm mesons in Fig.~1. For this the production rates of the 
states with the same or very close masses must be set according to 
simple total spin counting\footnote{\scriptsize Total spin counting 
works for the vector and pseudoscalar \bv\ and \bl\ 
\cite{Abvbt,Dbv,Lbv,Obv} or $\dl^{*0}_2$ and $\dl^0_1$ \cite{Odt}, 
although even in this case its small violation due to non 
negligible mass difference can not be excluded \cite{h_2}.}. 
For the $j_q =3/2$ and $j_q = 1/2$ states, with presumably different 
masses, this simple spin counting is expected to be violated. This 
violation can be accounted for assuming that the mass dependence of 
the production rates is described by the exponential with the same 
slope parameter $b$ as given earlier for the light-flavour mesons. 
Then the coefficient characterising the violation of simple spin 
counting for the \boz\ and \bt\ is
\beq
\varepsilon = 5\boz/(3\bt) = \mathrm{e}^{ b(M_{\bt} -M_{\boz}) }, 
\eeq
and the relative production rates of the four different states 
contributing to the $\bzz_{u,d}$ signal are set according to 
the following ``modified'' total spin counting 
procedure\footnote{\scriptsize With the $\dl^{\pm}$ and \dvp\ 
masses and the same $b$, Eq.~(8) yields $\varepsilon = 3\dl/\dv 
= 1.80 \pm 0.16$. This is consistent with $\varepsilon = 2.0 \pm 
0.3$ following from the value of $P_V=V/(P+V) = 0.595 \pm 0.045$ 
for these mesons \cite{Advp}, supporting our assumption that the 
violation of the simple spin counting rule is closely related to 
the mass difference.}:
\beq
         \bz:\boz:\bo:\bt  = \varepsilon:3\varepsilon:3:5.
\eeq
The values of $\varepsilon$ for the \bt\ and \boz\ masses determined
by L3 and OPAL are given in Table~3.  The \bz, \boz, \bo\ and \bt\ 
relative production rates in $b$-quark jets (with their overall rate 
given in Eq.~(7)) following from the modified total spin counting
(MTSC) rule proposed here are compared with those obtained using the 
simple total spin counting (STSC) applied by L3 and the proportion 
\bz:\boz:\bo:\bt = 2:2:3:3 used by OPAL.
\renewcommand{\arraystretch}{1.}
\begin{table}[htp]
\caption[] {\scriptsize The coefficient $\varepsilon$, relative
fractions of the four $P$-wave states and promptly produced 
$\bv(u,d)$ in $b$-quark jets (in \%) calculated with the \bt\ and 
\boz\ masses determined by L3 and OPAL and applying modified 
total spin counting (MTSC), simple total spin counting (STSC), and 
the OPAL proportion \bz:\boz:\bo:\bt  = 2:2:3:3.}  
\begin{center}
\begin{tabular}{|ccccc|} \hline
Experiment        &L3 (MTSC) &L3 (STSC) &OPAL (MTSC) &OPAL (2:2:3:3)\\
\hline
$\varepsilon$     &$1.50\pm 0.12$ &1   &$0.64\pm ^{0.10}_{0.12}$ & \\
\hline
$\bz(u,d)/\sigma_{b-\mathrm{jet}}$  &$2.75\pm 0.38$  &$2.15\pm 0.23$
                           &$1.56\pm 0.34$  &$5.2\pm 0.5$\\
$\boz(u,d)/\sigma_{b-\mathrm{jet}}$ &$8.3 \pm 1.1$   &$6.5\pm 0.7$
                           &$4.7 \pm 1.0$   &$5.2\pm 0.5$\\
$\bo(u,d)/\sigma_{b-\mathrm{jet}}$  &$5.5 \pm 0.6$   &$6.5\pm 0.7$
                           &$7.3 \pm 0.8$   &$7.7\pm 0.8$\\
$\bt(u,d)/\sigma_{b-\mathrm{jet}}$  &$9.2 \pm 1.0$   &$10.8\pm 1.1 $
                           &$12.2\pm 1.4$   &$7.7\pm 0.8$\\
$\bv(u,d)/\sigma_{b-\mathrm{jet}}$ 
                           &$39.6\pm 4.2$   &$39.7\pm 4.2$  
                           &$39.9\pm 4.3$   &$41.2\pm 4.5$\\ 
\hline
\end{tabular}
\end{center}
\end{table}

The relative \bv\ production rate in $b$-quark jet, 
$\sigma_{\bv}/\sigma_{b-\mathrm{jet}}$, was measured by the LEP 
experiments for a mixture of the states $\bl^*_d$, $\bl^*_u$ and 
$\bl^*_s$ with the following results: $0.677\pm 0.073$ \cite{Abvbt}, 
$0.650\pm 0.063$ \cite{Dbv}, $0.690\pm 0.086$ \cite{Lbv} and 
$0.660\pm 0.085$ \cite{Obv}, with the averaged value of $0.667\pm 
0.037$. Assuming that $0.3\bl^*_s$ are produced for each $\bv_d$, 
we obtain
\beq
     \sigma_{\bv(u,d)}/\sigma_{b-\mathrm{jet}} = 0.580 \pm 0.032.
\eeq
For determining the rate of the promptly produced $\bv(ud)$, the 
decays of $P$-wave mesons into \bv\pis\ have to be taken into
account. For this, the branching fractions into \bv\pis\ shown in 
Table~2 were used (the same as in \cite{Lbt,Abtexcl}). With the 
relative production rates from Table~3 calculated using modified 
total spin counting, this yields $Br(\bl_J \rightarrow \bv\pi) = 
0.714 \pm 0.029 \pm 0.068$ and $0.703 \pm 0.040 \pm 0.074$ for the 
L3 and OPAL results respectively. The additional systematic errors
account to half of the difference between these values and the 
value $Br(\bl_J \rightarrow \bv\pi(X)) = 0.85 \pm 0.29$ found by 
OPAL \cite{Ob1nar}. The resulting relative rates of the promptly 
produced $\bv(ud)$ are presented in Table~3, together with 
similarly obtained values based on the L3 results with simple 
total spin counting and OPAL results with the OPAL proportion of 
the relative rates.

As one can see from Table~3, the relative rates of the four spin
states, and especially the \bz/\bt\ ratio, are quite sensitive to 
the spin counting rules assumed. For the masses of these states 
from OPAL, this ratio obtained using the OPAL proportion of the 
rates is larger by a factor of 5.2 than the same ratio obtained 
with modified total spin counting. This suggests that significantly 
different fitted values of the masses might be obtained if modified 
total spin counting were applied instead of the OPAL proportion. 
On the other hand, the rate of promptly produced \bv\ is practically 
insensitive to the difference in the counting rules. The \bt\ rates 
obtained with modified spin counting at the masses from the L3 and 
OPAL are consistent within 1.7 standard deviations (or even less 
since the \bt\ mass from L3 is larger than from OPAL). This suggests 
that the production rates of  the \bt\ and promptly produced \bv\ 
are sufficiently reliable to allow comparison of their mass 
dependence with other data.

The relative rates of \boz, \bt\ and promptly produced \bv\ in 
$b$-quark jets calculated using modified total spin counting 
and divided by the spin counting factors $2J + 1$ are shown at the 
\boz\ and \bt\ masses from L3 in Fig.~1a and at the \boz\ and 
\bt\ masses from OPAL in Fig.~1b. The fits of the data to six 
exponentials with different normalization parameters for the 
six meson families, but the {\em same\/} slope parameter $b = 4.17 
\pm 0.21$ $(\mathrm{GeV}/c^2)^{-1}$ in Fig.~1a and $b = 4.01 \pm 
0.19$ $(\mathrm{GeV}/c^2)^{-1}$ in Fig.~1b describe the data well 
(solid lines in Fig.~1). The slope parameters are very close to 
the value $b = 4.11 \pm 0.27$ $(\mathrm{GeV}/c^2)^{-1}$ obtained 
for the light-flavour mesons. Thus we see that the mass 
dependences of the production rates per spin projection are 
indeed very similar for the light-flavour, charm and bottom 
mesons\footnote{\scriptsize For the bottom mesons, this applies, 
strictly speaking, only to the \bv\ and \bt, since for the \boz\ 
and \bt\ rates the same mass dependence as for the light-flavour 
mesons has been imposed by Eqs.~(8) and (9).}. One important lesson 
from this observation is the existence of a close relationship 
between the masses of the $P$-wave states and their production 
rates. If the masses of the $j_q = 1/2$ states with $J^P = 0^+$ 
and $1^+$ are below (above) the masses of the $j = 3/2$ states 
with $J^P = 1^+$ and $2^+$, their production rates per spin 
projection are larger (smaller) than for the $j_q = 3/2$ states, 
as shown for the \boz\ and \bt\ in  Fig.~1a (Fig.~1b).

Apart from the $P$-wave mesons discussed above, only the production 
rates of the scalars \azn\ and \fzn\ were  measured at LEP 
\cite{Drofzftkt,Ophfzft,Oazom}. They are presented in Table~1 and 
Fig.~1. The \azn\ rate is consistent within errors with the mass 
dependence of the \roz, \oms\ and \ftn\ rates. The \fzn\ rate is 
consistent with the \azn\ rate as expected, but appears to be 
slightly higher than follows from the mass dependence of the \roz, 
\oms\ and \ftn\ rates. However, this might well be due to 
overestimated fractions of the promptly produced \azn\ and \fzn, 
which are difficult to estimate. These presumably must be comparable 
with those for the vector mesons (with similar masses), but are 
higher in JETSET (see Table~1). Such an explanation is supported by 
the mass dependence of the \roz, \oms, \azn, \fzn\ and \ftn\ total 
rates \cite{Uvarov}. If the production rates of other $P$-wave mesons 
follow the same mass dependence as observed in Fig.~1, this allows 
their production rates to be estimated. For example, the 
corresponding predictions for the $b_1(1235)$ and $f_1(1420)$ total 
production rates per \zo\ hadronic decay are $0.102\pm 0.031$ and 
$0.0126 \pm 0.0045$ if the $f_1(1420)$ is a pure $s\bar s$ state. 

Moreover, provided that the observed mass dependence of production 
rates is indeed universal for all flavours, it allows not only the
production rates of mesons with different flavours to be related, 
but also their masses. Indeed, from simple mass rescaling in Fig.~1 
one obtains the following phenomenological mass formulae:   
\beq
  \bl_i = \bt - (\bt - \bv) {{T-P_i}\over{T -V}},~~~
  \dl_i = \dt - (\dt - \dv) {{T-P_i}\over{T -V}}
\eeq
where $V$, $T$ and $P_i$ are the masses of the vector, tensor 
and $P$-wave (with $J^P = 1^+$ or $0^+$) light-flavour mesons 
corresponding to the masses of their respective charm \dv, \dt\  
and $\dl_i$, and bottom \bv, \bt\ and $\bl_i$ partners. 

From Eq.~(11), with the \kvo, \kto, \kah, \dv, \dt\ and 
\bv\ masses from PDG \cite{PDG} and the \bt\ mass, $5752 \pm 
15$~MeV/$c^2$, taken as the average of the masses obtained by ALEPH, 
L3 and OPAL (Table 2) and with the error equal to half of the 
difference between the ALEPH and L3 values, one obtains:
\beq 
   M_{\bo}     = 5728 \pm 16~{\mathrm{MeV}}/c^2,~~~ 
   M_{\dl^0_1} = 2433 \pm 6~{\mathrm{MeV}}/c^2. 
\eeq
With the \kal\ instead of the \kah\ in Eq.~(11) one has:
\beq 
     M_{\boz} =       5625 \pm 16~{\mathrm{MeV}}/c^2,~~~
     M_{\dl^{*0}_1} = 2325 \pm 6~{\mathrm{MeV}}/c^2.
\eeq
The limited accuracy of the phenomenological formulae (11) results
in additional systematic uncertainty, not accounted for in the 
mass estimates given in Eqs.~(12) and (13). It can roughly 
be estimated from the mass relation $(\bt - \bl_i)/(\bt - \bv)
= (\dt -\dl_i)/(\dt - \dv)$ following from Eq.~(11), which
imposes practically the same mass splitting between the \bt\
and \bo\ as between the \dt\ and $\dl_1$. This is not consistent 
with the smaller \bt\ and \bo\ mass difference of 12 MeV/$c^2$,
required in the fits performed by the LEP experiments, in
comparison with the measured \dto\ and $\dl^0_1$ mass difference
$37\pm 3$ MeV/$c^2$ and may result in possible biases of
$\approx 25$ MeV/$c^2$ in our mass estimates.

The \bo\ mass given in Eq.~(12) agrees within errors with 
the averaged value of this mass $5740 \pm 15$ MeV/$c^2$ from 
ALEPH, L3 and OPAL and $M_{\bo} = 5710 \pm 20 $ MeV/$c^2$ from CDF. 
The mass difference $M_{\bt} - M_{\bo} = 24 \pm 22$ MeV/$c^2$ is 
consistent with the constraint of 12 MeV/$c^2$ imposed by the 
LEP experiments. The \boz\ mass given in Eq.~(13) is consistent 
within 2 standard deviations with $M_{\boz} = 5670 \pm 16$ MeV/$c^2$ 
from the L3 fit, agrees with $M_{\boz} = 5639 \pm ^{10}_{12}$ 
MeV/$c^2$ from the ALEPH fit, but significantly smaller than the 
value following from the OPAL fit. The obtained $\dl^0_1$ mass is 
in good agreement with the PDG value $2422.2 \pm 1.8$ MeV/$c^2$. 
The $\dl^{*0}_1$ mass in Eq.~(13) represents our prediction for 
the mass of the broad, not yet established state. 

The physical \kal\ and \kah\ are mixtures of the two SU(3) octet 
states $^1P_1$ and $^3P_1$. The decay patterns of the \kal\ and
\kah\ suggest that these singlet and octet states are almost
degenerate, with a mixing angle near $45^{\circ}$. Thus, from 
the decay amplitudes of the \kal\ and \kah\ into $\rho\kl$ and 
$\kl^*\pi$, the ACCMOR collaboration found $\theta = 56^{\circ}
\pm 3^{\circ}$ \cite{Daum}. Provided that the heavy quark limit 
is also appropriate for the strange mesons, the two mixed \ka\ 
mass eigenstates of $J^P =1^+$ can also be described by the 
total angular momentum $j_q$ of the light quark with $j_q = 1/2$ 
and $j_q = 3/2$ expected to be degenerate with the $J^P = 0^+$ 
and $J^P = 2^+$ states, respectively. By a change of basis one 
can introduce a new mixing angle $\theta_K$ which defines the 
amount of $j_q = 1/2$ and $j_q = 3/2$ in the physical \kal\ 
and \kah\ states. According to Isgur \cite{Isgur} (see also 
\cite{Eichten,Gubank,Blundell}), the \kal\ and \kah\ are quite 
near to being the pure $j_q = 3/2$ and $j_q = 1/2$ states, 
respectively. The small splitting between \kzn\ and \kah, 
implying that the $0^+(1/2)$ and $1^+(1/2)$ levels are nearly 
degenerate as expected for all heavy-light systems, is well 
consistent with such association. However, this is certainly 
not a case for the \ktn\ and \kal\ associated with the $2^+(3/2)$ 
and $1^+(3/2)$ levels, respectively. 

On the other hand, our results following from the mass relations
(11) suggest that the \kah\ and \kal\ might be assigned in the 
heavy quark limit to the $j_q = 3/2$ and $j_q = 1/2$ levels, 
respectively. This may certainly imply either that our assumption 
about a universal mass dependence of the production rates for all 
$P$-wave states (including the \kah\ and \kal) fails, or that 
our phenomenological mass formulae (11) resulting from this 
assumption are not correct. However, if it is not the case, 
such an assignment implies that the $j_q = 3/2$ levels 
corresponding to the \kah\ and \ktn\ are degenerate, whereas 
the $j_q = 1/2$ levels corresponding to the \kal\ and \kzn\ 
are quite different, just contrary to the situation discussed 
earlier. We also notice that an attempt to apply Eq.~(11) 
with the \kzn\ mass for the determination of the \bz\ and 
$\dl_0$ masses results, due to the small difference between 
the \ktn\ and \kzn\ masses, in $M_{\bz} \approx M_{\bt}$ and 
$M_{\dl_0} \approx M_{\dt}$. This is not consistent with the 
values of the \boz\ and $\dl^{*0}_1$ masses given in Eq.~(13), 
if the mass difference between the \boz\ and \bz, and also the 
$\dl^{*0}_1$ and $\dl_0$, is small as expected. For the \bz\ and 
\dz\ masses equal to the \boz\ and $\dl^{*0}_1$ masses given in 
Eq.~(13), Eq.~(11) by definition gives the $\kl^*_0$ mass equal 
to the \kal\ mass. A smaller $\kl^*_0$ mass is also required in 
the description of the light-flavour $P$-wave mesons in the 
nonrelativistic quark model \cite{h_3}. As noticed in \cite{h_3}, 
this can be explained if the observable \kzn\ mass is replaced 
by its ``bare'' $q\bar q$ mass corresponding to the $K$-matrix 
pole. In the $K$-matrix analysis of the $0^{++}$-wave \cite{anis}, 
the ``bare'' $\kl^*_0$ mass, in one of the two possible solutions, 
is $1220 \pm 70$ MeV/$c^2$, consistent with the \kal\ mass. Thus, 
if this conjecture is correct, the \kah\ and \kal\ assignment in 
the heavy quark limit to the $j_q = 3/2$ and $j_q = 1/2$ levels 
is consistent with the expected degeneracy of the $1^+(3/2)$ and 
$2^+(3/2)$ and, respectively, the $1^+(1/2)$ and $0^+(1/2)$ levels. 
It also provides a consistent description of the strange, charm 
and bottom meson production rates and also their masses and lends 
support to the models suggesting that the $j_q = 1/2$ levels 
for the strange, charm and bottom mesons are {\em below\/} the 
$j_q =3/2$ levels. In particular, our results given in Eq.~(13) 
are in excellent agreement with the predictions \cite{Gronau}. 

In conclusion we have shown that the mass dependences of the  
production rates for the six families of primary produced mesons in 
\zo\ hadronic decays obtained from results of the LEP experiments: 
the vector and tensor light-flavour mesons, the vector and $P$-wave 
charm, strange charm and bottom mesons are very similar. This allows 
not only the production rates of mesons with different flavours to 
be related, but also their masses, thus showing an interesting 
connection between hadron production properties and their masses. 
Our analysis suggests that the $0^+(1/2)$ and $1^+(1/2)$ levels 
are {\em below\/} the $1^+(3/2)$ and $2^+(3/2)$ levels not only for 
the charm and bottom but also for the strange mesons. Contrary to 
the conventional picture, the strange axial mesons \kal\ and \kah\ 
might be considered as mainly $1^+(1/2)$ and $1^+(3/2)$ levels, 
respectively, degenerate with the $0^+(1/2)$ and $2^+(3/2)$ levels 
of the \kzn\ and \ktn\, if the observed \kzn\ mass is replaced by 
its ``bare'' $q\bar q$ mass corresponding to the $K$-matrix pole 
and close to the \kal\ mass. Although these results, if verified 
by future experiments, do not support the spin-orbit inversion 
suggested by Isgur \cite{Isgur}, they amusingly lend strong support 
to his conclusion about the key role that the strange quark plays 
as the link between heavy- and light-quark hadrons. 

\section*{Acknowledgements}
I thank S. Godfrey and W. Venus for discussions and helpful comments.

\end{document}